\documentstyle[12pt,epsfig]{article}
 \newcommand{\p}{{\bf p}}

 \newcommand{\bK}{{\bf K}}
 \newcommand{\0}{{\bf 0}}

 \newcommand{\lam}{{\lambda}}
 \newcommand{\ra}{{\rangle}}
 \newcommand{\la}{{\langle}}
 \newcommand{\be}{\begin{equation}}
 \newcommand{\ben}{\begin{eqnarray}}
 \newcommand{\een}{\end{eqnarray}}
 \newcommand{\ee}{\end{equation}}
 
 \newcommand{\po}{\stackrel{o}{\p}}

\begin{document}

\begin{center}
{\bf 
Axial and Vector Coupling Constants of Nucleons
and Constituent Quarks \\ 
in the Quasipotential Model}
\\
 T.P.Ilichova
\\[6pt] {\it Francisk
Skaryna Gomel State University, LPP JINR, Dubna} \\
ilicheva@cv.jinr.ru
\end{center}

\begin{abstract}
\par
Calculation of ratio of the nucleon axial
and vector coupling constants $G_A/G_V$  in the framework
of the quasipotential quark model has been performed. 
 $G_A/G_V$ is defined by the ratio of the
corresponding quark axial and and vector coupling
constants $g_A/g_V$ and by the dimensionless parameter 
describing the relativistic relative quark motion
and vanishing in the nonrelativistic limit.
  This relativistic parameter also defines
nucleon magnetic moments.
  Thus we obtain relations independent of the form of
the wave function:
$G_A/G_V=(g_A/g_V)5(\mu^p-1)/6$, $\mu^p=(1-4\mu^n)/3$.
Proton magnetic moment $\mu^p$ from the second relation  
reproduces the experimental
value with the error of 3 $\%$. 
  Then the first relation allows us
to find $(g_A/g_V)$ with 5\% 
uncertainty: $(g_A/g_V)\in[0.809,0.850]$.
\end{abstract}

\par
The work is devoted to the calculation of ratio
of the axial and vector coupling constant $G_A/G_V$ for nucleon
in the quasipotential quark
model~\cite{IzvVuzov_Ilicheva1,IzvVuzov_Ilicheva2}.
In the nonrelativistic quark model
$G_A = 3\la p|\sigma^{(3)}_3 t^{(3)}_+|n\ra g_A = 5/3 g_A$,
$G_V = 3\la p|t^{(3)}_+|n\ra g_V=g_V$,
where $t^{(3)}_+$ is the isospin increasing operator
and $\sigma^{(3)}_3$ is the Pauli spin matrix
acting to the third quark in the $SU(6)$ nucleon wave function
for proton $|p\ra$ and neutron $|n\ra$;
$g_{A/V}$ is the quark axial/vector coupling constants.
Experimental value
$G_A/G_V = 1.2695\pm0.029$~\cite{ParticleDataGroup2004}
yields
$g_A/g_V = 0.762$ in the nonrelativistic quark model.
\par
As a rule, relativistic models have at least 3 dimensional parameters:
nucleon mass $M$, quark mass $m$ and the parameter which characterizes
the interaction in the system (we sign it as $\gamma$).
In this case the dimensionless nucleon magnetic moments
(in units of nuclear magneton) are defined by
the ratio of the nucleon mass and one or more relative quark motion
parameters.
In the relativistic constituent quark model $G_A/G_V$,
$\mu^p$ and $\mu^n$ are related.
For example, in light-front
models~\cite{dziembovski2,brodsky_schlumpf,chung_coester}
the nucleon magnetic moments depend on $m/M$ and on $m/\gamma$.
For this case $G_A/G_V$ depends on the relative quark
motion parameters $m/\gamma$ only.
In the quark bag model (see.~\cite{close} and references there in)
the nucleon magnetic moments are defined by the dimensionless ratio
of the nucleon mass and bag radius ($\mu^p\approx 0.4RM$).
In this case $G_A/G_V$ is defined by
the other dimensionless relation $mR$ ($m$ is the quark mass).
\par
Described in works~\cite{IzvVuzov_Ilicheva1,IzvVuzov_Ilicheva2}
quasipotential quark model
has the following feature:
nucleon magnetic moments  depend on
the dimensionless parameter defined only by parameters of
the relative quark motion ($m/\gamma$) and
don't depend on the nucleon mass.
In work~\cite{IzvVuzov_Ilicheva2} it was shown that
relativistic kinematics is a only factor which influences
the parametric dependence of the nucleon magnetic moments.
\par
Let us consider $G_A/G_V$
in model~\cite{IzvVuzov_Ilicheva1, IzvVuzov_Ilicheva2}.
In the rest frame of the initial
nucleon the electroweak current has the form:
\be
J_\mu^{w\lam_J\lam'_J}(\bK,\0)\equiv
\la E^M_{\bK} \bK \lam'_J;p
\mid J^w_{\mu}(0)\mid M\0 \lam_J;n\ra =
\bar U^{\lam'_J}_{p}(\bK)
[G_V(t)\gamma_{\mu}
+
G_A(t)\gamma_{\mu}\gamma_5
]T_+U^{\lam_J}_{n}(\0),
\label{electroweak_mu_current_nuclon}
\ee
where
$T_+=T_1+iT_2$ ($T_1$, $T_2$ are components of
the nucleon isospin operator).
We assume spin projections of the initial and final
nucleons to be equal to
$\lam'_J=\lam_J=1/2$ and further we will omit this indices.
For zero momentum transfer we have:
\be
G_V\equiv G_V(0)=J_{0}^{w}(\0,\0),
\quad
G_A\equiv G_A(0)= {\bf J}^{w}_{3}
 (\0,\0) .
\label{GAV_def}
\ee
\par
In the three quark quasipotential model the electroweak current
$J_{\mu}^{w}(\bK,{\0})$
 has the form
coinciding with the expression
for electromagnetic current~\cite{IzvVuzov_Ilicheva2}
(notations as in~\cite{IzvVuzov_Ilicheva2}):
\be
 J_{\mu}^{w}(\bK,{\0}) =
 3 \int d\Omega_{\p_1} d\Omega_{\p_2}
 \varphi(\po_1,\po_2,\po'_3)
(\chi_p M^{(1)} M^{(2)} \Gamma^{(3)}_{\mu} \chi_n)
\varphi(\p_1,\p_2,\p_3)/E_{\p_3}E_{\po'_3} ,
\label{current}
\ee
where matrices $M^{(1)}, M^{(2)}, \Gamma^{(3)}_{\mu}$
are defined in~\cite{IzvVuzov_Ilicheva2}
with the replacement of the electromagnetic quark current
to electroweak quark current:
$j^{(3)}_{\mu}(\p'_3,\p_3)
\to j_{w\mu}^{(3)}(\p'_3,\p_3)=
\bar u(\p'_3)
[
g_V(t)\gamma_{\mu}
+
g_A(t)\gamma_{\mu}\gamma_5
]
t_+^{(3)}u(\p_3).
$
We rewrite the electroweak quark current in the two-spinors form:
\ben
&&j_{0}^{(3)w}(\p'_3,\p_3)
=
\frac{1}{2\sqrt{(m+E_{\p'_3})(m+E_{\p_3})}}
\{g_V
[(m+E_{\p'_3})(m+E_{\p_3})+
(\sigma^{(3)}\p'_3)
(\sigma^{(3)}\p_3)]
+
\nonumber\\
&&
+g_A
[(\sigma^{(3)}\p_3)(m+E_{\p'_3})+
(\sigma^{(3)}\p'_3)(m+E_{\p_3})
]\}t_+^{(3)};
\label{electroweak_0_current_quark}
\een
\ben
&&(\varepsilon_{(3)} {\bf j}^{(3)w}(\p'_3,\p_3))
=
\frac{1}{2\sqrt{(m+E_{\p'_3})(m+E_{\p_3})}}*
\nonumber\\
&&
*
\{
g_V
[(m+E_{\p'_3})
(\sigma^{(3)}\varepsilon_{(3)})
(\sigma^{(3)}\p_3)
+(m+E_{\p_3})
(\sigma^{(3)}\p'_3)
(\sigma^{(3)}\varepsilon_{(3)}]
+
\nonumber\\
&&
+g_A
[(m+E_{\p'_3})
(\sigma^{(3)}\varepsilon_{(3)})(m+E_{\p_3})
+
(\sigma^{(3)}\p'_3)
(\sigma^{(3)}\varepsilon_{(3)})
(\sigma^{(3)}\p_3)]
\}t_+^{(3)}.
\label{electroweak_3_current_quark}
\een
Taking into acount the wave function
normalization condition (see~\cite{IzvVuzov_Ilicheva2}),
from(\ref{current})
we have:
\be
G_V=3g_V
\la p| t^{(3)}_{+} |n\ra=g_V,\quad
G_A=
g_A
\la p|\sigma^{(3)}_3 t^{3}_{+} |n\ra
 (3-2a)
\label{G_A_V}
\ee
\be
a=1-
\int d\Omega_{\p_1} d\Omega_{\p_2}
|\varphi(\p_1,\p_2,\p_3)|^2
m/E^2_{\p_3}>0.
\label{a_approx}
\ee
The last expressions give us the ratio:
\be
\frac{G_A}{G_V}=
\frac{g_A}{g_V}(\frac{5}{3}-\frac{10}{9}a).
\ee
The parameter $a$ defines the nucleon magnetic
moments also:
$\mu^p=3-4a/3$, $\mu^n=-2+a$~\cite{IzvVuzov_Ilicheva2}.
If we fix $a=0.121$ ($m/\gamma=1.82$)
(for agreement of the nucleon
magnetic moments with the experimental data
with errors 2\%~\cite{IzvVuzov_Ilicheva2}),
our prediction is consistent
with the $G_A/G_V$ experimental value for
$g_A/g_V =0.827$.
\par
From~(\ref{G_A_V}) and from formulae for $\mu^p$, $\mu^n$
 we have the following relations for $G_A/G_V$, $\mu^p$
and $\mu^n$ independent of the wave function form:
\be
\mu^p=\frac{(1-4\mu^n)}{3},\quad
\frac{G_A}{G_V}=
\frac{g_A}{g_V}\frac{5(\mu^p-1)}
{6}.
\label{G_A_G_V_cherz_mu}
\ee
From the first relations
we have obtained that value $\mu^p$ is consistent
with the exparimental values with the error of 3.3\%.
The second relation allows us to estimate the quark $g_A/g_V$
with uncertainty of 5\%: $(g_A/g_V)\in[0.809,0.850]$.
The lower and upper limits were obtained by substituting
the experimental values of the neutron and proton
magnetic moments, respectively,
in the second relation~(\ref{G_A_G_V_cherz_mu}).

\par
Thus, in this model the nucleon magnetic moments and $G_A/G_V$
depend on the one model parameter and are related by the
expression~(\ref{G_A_G_V_cherz_mu}), which does not depend
on the form of the relative motion wave function.
$G_A/G_V$ defines unique quark ratio $g_A/g_V$.

\end{document}